\documentclass[runningheads,a4paper]{else}

\usepackage{amssymb,amsmath}
\usepackage{graphicx}
\usepackage{color}

\newtheorem{theo}{Theorem}
\newtheorem{lemm}[theo]{Lemma}
\newtheorem{coro}[theo]{Corollary}

\newtheorem{prob}[theo]{Problem}
\newtheorem{defi}[theo]{Definition}

\begin{document}

\begin{frontmatter}



\author[tubs]{S{\'a}ndor P.\ Fekete}
\ead{s.fekete@tu-bs.de}
\author[tubs]{Nils Schweer}
\ead{n.schweer@tu-bs.de}
\author[tubs]{Jan-Marc Reinhardt}
\ead{j-m.reinhardt@tu-bs.de}

\address[tubs]{Department of Computer Science, TU Braunschweig, Germany} 

\title{A Competitive Strategy for \\ Distance-Aware Online Shape Allocation}

\begin{abstract}
We consider the following online allocation problem:
Given a unit square $S$, and a sequence of numbers $n_i\in \{0,1\}$ 
with $\sum_{j=0}^i n_j\leq 1$; at each step $i$, select a region $C_i$ of
previously unassigned area $n_i$ in $S$. The objective is to make these regions
compact in a distance-aware sense:
minimize the maximum (normalized) average Manhattan distance between points from
the same set $C_i$. Related location problems have received a considerable amount of attention;
in particular, the problem of determining the ``optimal shape of a city'', i.e.,
allocating a {\em single} $n_i$ has been studied.
We present an online strategy,
based on an analysis of space-filling curves; for continuous shapes,
we prove a factor of 1.8092, and 1.7848 for discrete point sets.
\end{abstract}

\end{frontmatter}


\section{Introduction}
Many optimization problems deal with allocating point sets to a given
environment. Frequently, the problem is to find compact allocations,
placing points from the same set closely together. One well-established
measure is the average $L_1$ distance between points.
A practical example occurs in the context of grid computing, where one
needs to assign a sequence of jobs $i$, each requiring an (appropriately
normalized) number $n_i$ of processors,
to a subset $C_i$ of nodes of a large square grid, such that the average 
communication delay between nodes of the same job is minimized;
this delay corresponds to the number of grid hops \cite{leung},
so the task amounts to finding subsets with a small 
average $L_1$, i.e., {\em Manhattan} distance. 
Karp et al.~\cite{karp} studied the same problem in the context of memory allocation.

\begin{figure}
  \hfill{\includegraphics[width=.3\textwidth]{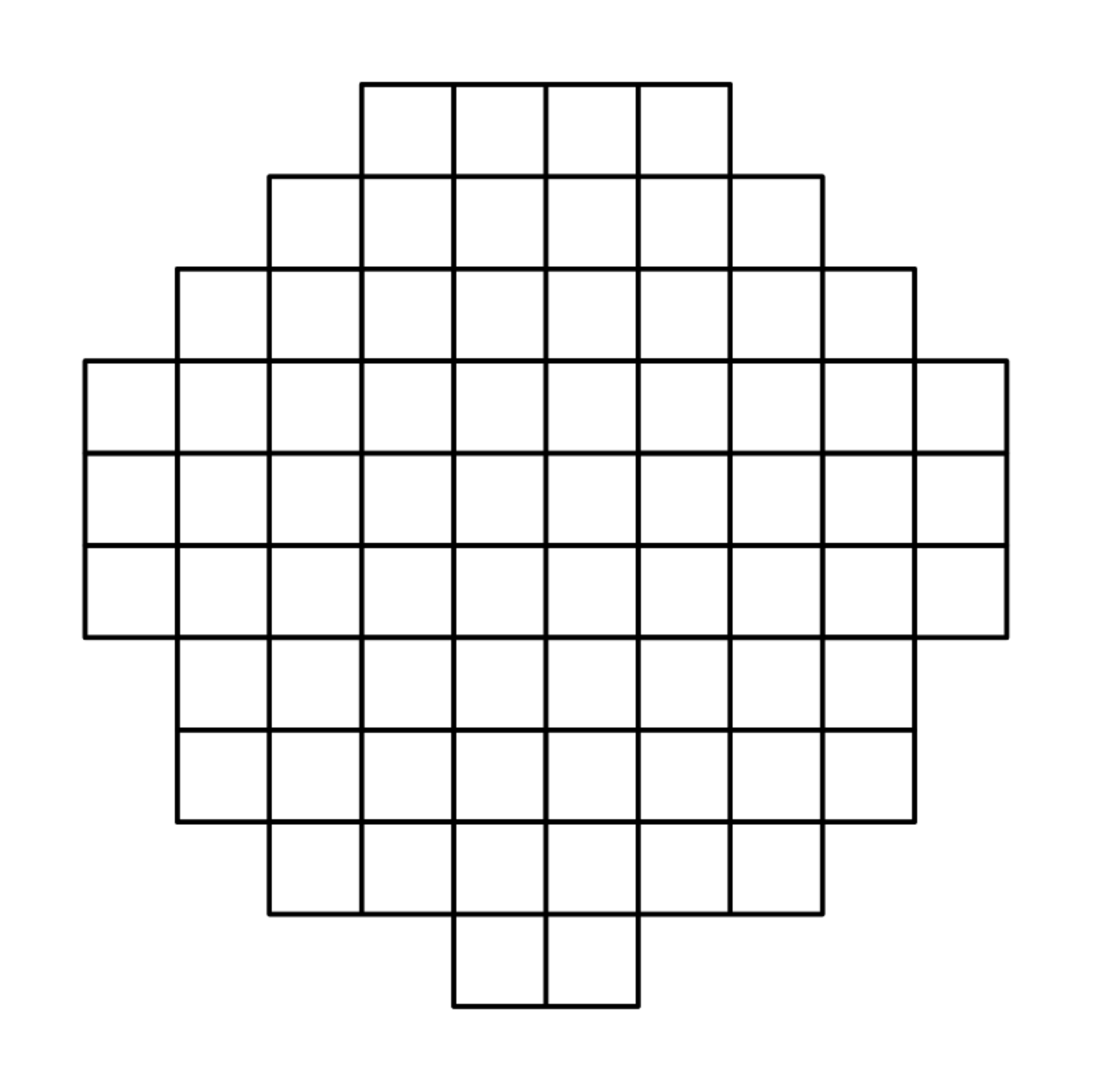}}\hfill
  {\includegraphics[width=.5\textwidth]{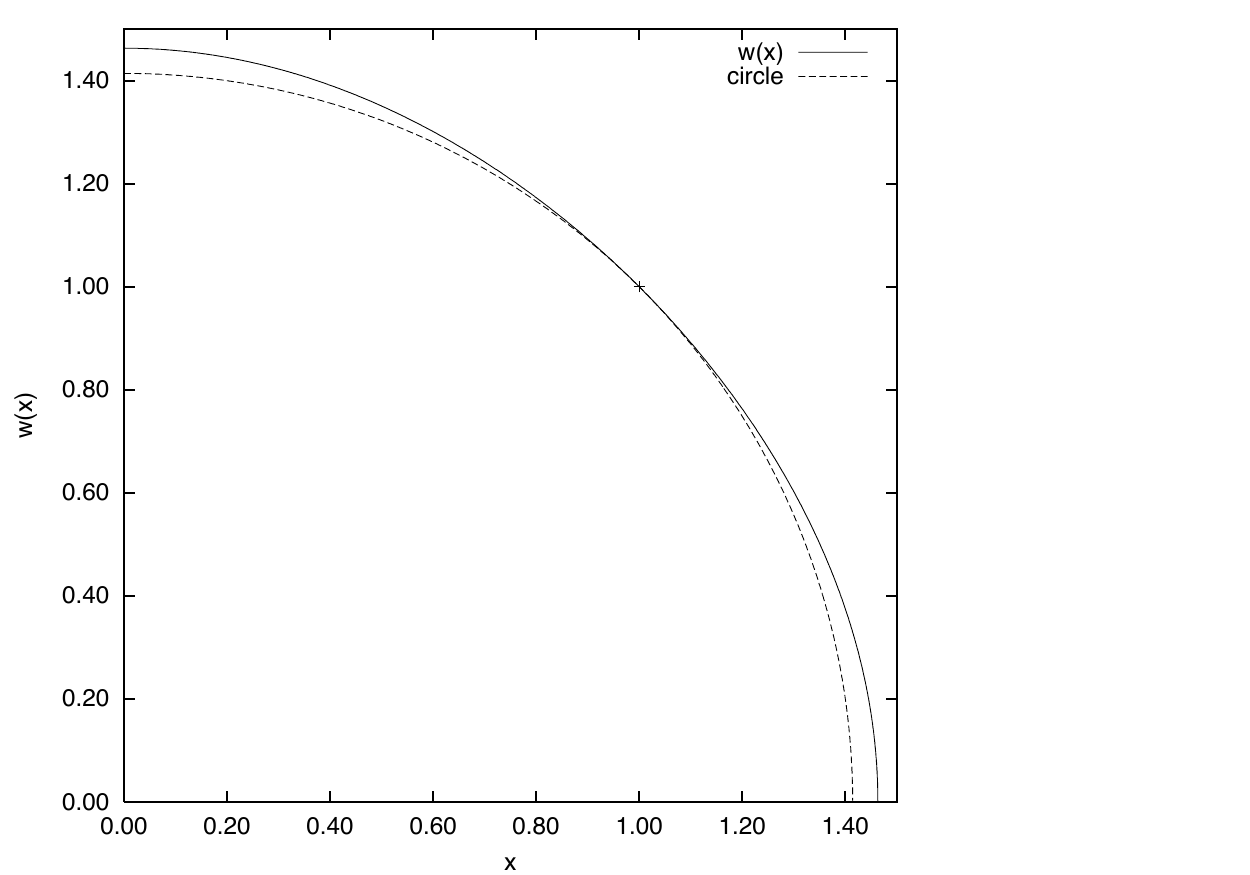}}\hfill
  \caption{Finding optimal individual shapes. (Left) An optimal shape composed of n=72 grid cells, according to~\cite{dfr-psmd-11}.
(Right) The optimal limit curve $w(x)$, according to~\cite{bender2}.}\label{fig-related}
\end{figure}

Even in an offline setting without occupied nodes, 
finding an optimal allocation for one set of size $n_i$ is
not an easy task; as shown in Fig.~\ref{fig-related}, the results are typically ``round'' shapes.
If a whole sequence of sets have to be allocated, packing such 
shapes onto the grid will produce gaps, causing later 
sets to become disconnected, and thus leads to extremely bad average distances.
Even restricting the shapes to be rectangular is not a remedy, as the 
resulting problem of deciding whether a set of squares 
(which are minimal with respect to $L_1$ average distance among all rectangles)
can be packed into a given square container 
is NP-hard~\cite{squaresquare}; moreover,
disconnected allocations may still occur.

In this paper, we give a first algorithmic analysis for
the {\em online} problem. Using an allocation scheme based
on a space-filling curve, we establish competitive factors of 
1.8092 and 1.7848 for minimizing the maximum average Manhattan distance 
within an allocated set.

\subsection*{Related Work}
Compact location problems have received a considerable amount of attention.
Krumke et al.~\cite{krumke} have considered the {\em offline} 
problem of choosing a set of $n$ vertices in
a weighted graph, such that the average distance is minimized.
They showed that the problem is NP-hard (even to approximate); for the scenario in which 
distances satisfy the triangle inequality, they gave algorithms that achieve asymptotic approximation factors of 2.
For points in two-dimensional space and Manhattan distances, Bender et al.~\cite{bender2} gave a simple 1.75-approximation
algorithm, and a polynomial-time approximation scheme for any fixed dimension.

The problem of finding the ``optimal shape of a city'', i.e., a shape
of given area that minimizes the average Manhattan distance, was first considered
by Karp, McKellar, and Wong \cite{karp}; independently, Bender, Bender,
Demaine, and Fekete \cite{bender} showed that this shape can be characterized
by a differential equation for which no closed form is known.
For the case of a finite set of $n$ points that needs to be allocated to a grid,
Demaine et al.~\cite{dfr-psmd-11} showed that there is an 
$O(n^{7.5})$ dynamic-programming algorithm, which allowed them to compute all optimal shapes
up to $n=80$. Note that all these results are strictly offline, even though
the original motivation (register or processor allocation) is online.

Space-filling curves
for processor allocation with our objective function have been used before, see Leung et al.~\cite{leung}; 
however, no algorithmic results and no competitive factor was proven.
Wattenberg~\cite{w-nsfvs-05} proposed an allocation scheme similar to ours,
for purposes of minimizing the maximum {\em Euclidean diameter} of an allocated shape.
Like other authors before (in particular, Niedermeier et
al.~\cite{nrs-tolmi-02} and Gotsman and Lindenbaum~\cite{gl-mpdsf-96}), he
considered a measure called $c${\em -locality}:
for a sequence $1,\ldots, i, \ldots, j, \ldots$ of points on a line, a
space-filling mapping $h(.)$ will guarantee $L_2(h(i),h(j))<c\sqrt{|j-i|}$, for
a constant $c$ that is $\sqrt{6}\approx 2.449$ for the Hilbert curve, and 2 for the
so-called H-curve. One can use $c$-locality for establishing a
constant competitive factor for our problems; however, given that 
the focus is on bounding the worst-case distance ratio for an embedding instead
of the average distance, it should come as no surprise that the resulting
values are significantly worse than the ones we achieve. 
On a different note, de Berg, Speckmann, and van der Weele ~\cite{DBLP:journals/corr/abs-1012-1749}
consider treemaps with bounded aspect ratio. Other related work includes Dai and Su~\cite{ds-lpsfc-03}.

\subsection*{Our Results}
We give a first competitive analysis for the online shape allocation problem within a given bounding box, with the objective
of minimizing the maximum average Manhattan distance. In particular, we give the following results.

\begin{itemize}
\item We show that for the case of continuous
shapes (in which numbers $n_i$ correspond to area), a strategy based on a space-filling Hilbert curve
achieves a competitive ratio of 1.8092. 
\item For the case of discrete point sets (in which numbers 
indicate the number of points that have to be chosen from an
appropriate $N\times N$ orthogonal grid), we prove a competitive factor
of 1.7848. 
\item We sketch how these factors may be further improved,
but point out that a Hilbert-based strategy is no better than a competitive
factor of 1.3504, even with an improved analysis.
\item We establish a lower bound of 1.144866 for {\em any} online strategy in the case of discrete point sets, and argue the existence of a lower bound for the continuous case.
\end{itemize}

The rest of this paper is organized as follows. In Section~2, we give some basic
definitions and fundamental facts. In Section~3, we provide a brief description
of an allocation scheme based on a space-filling curve. Section~4 gives a mathematical
study for the case of continuous allocations, proving
that the analysis can be reduced to a limited number of shapes, and establishes
a competitive factor of 1.8092. Section~5 sketches a similar analysis for the case
of discrete allocations; as a result, we prove a competitive factor of 1.7848.
Section~6 discusses lower bounds for online strategies.
Final conclusions are presented in Section~7.

\section{Preliminaries}
We examine the problem of selecting shapes from a square, such that
the maximum average $L_1$-distance of the shapes is minimized. 
We first formulate the problem more precisely. This covers both the
continuous and the discrete case; the former arises as the limiting case
of the latter, while the latter needs to be considered for
allocations within a grid of limited size.

\begin{defi}
A \emph{city} is a (continuous) shape in the plane with fixed area. For a city $C$ of
area $n$, we call
\begin{equation}
c(C) = \frac{1}{2}\iiiint_{(x,y),(u,v) \in C} (|x-u|+|y-v|) \,\mathrm{d} v \,\mathrm{d} u \,\mathrm{d} y \,\mathrm{d} x
\end{equation}
the \emph{total Manhattan distance between all pairs of points in $C$} and
\begin{equation}
\phi(C) = \frac{2 \, c(C)}{n^{5/2}}
\end{equation}
the \emph{$\phi$-value} or \emph{average distance} of $C$. 
An $n$-\emph{town} $T$ is a subset of $n$ points in the integer grid. Its {\em normalized average Manhattan distance} is
\begin{equation}
\phi(T) = \frac{2 c(T)}{n^{5/2}} = \frac{\sum_{s\in T}\sum_{t\in T} \|s-t\|_1}{n^{5/2}}
\end{equation}
\end{defi}

The normalization with $n^{2.5}$ yields a dimensionless measure 
that remains unchanged under scaling, and makes the continuous
and the discrete case comparable; see \cite{bender}.

\begin{prob}
In the continuous setting, we are given a sequence $n_1, n_2, \ldots, n_k \in \mathbb{R}^+$
with $\sum_{i=1}^{k}{n_i} \le 1$. Cities $C_1, C_2, \ldots, C_k$ of size $n_1, n_2, \ldots, n_k$ are to be
chosen from the unit square, such that $\max_{1 \le i \le k}{\phi(C_i)}$
is minimized.

In the discrete setting, we are given a sequence $n_1, n_2, \ldots, n_k \in \mathbb{N}^+$
with $\sum_{i=1}^{k}{n_i} \le N^2$. Towns $C_1, C_2, \ldots, C_k$ of size $n_1, n_2, \ldots, n_k$ are to be
chosen from the $N\times N$ grid, such that $\max_{1 \le i \le k}{\phi(C_i)}$ is minimized.
\end{prob}

Although it has not been formally proven, the offline problem is conjectured to be NP-hard, see
\cite{schweer}; if we restrict city shapes to be rectangles, there is
an immediate reduction from deciding whether a set of squares can be packed
into a larger square \cite{squaresquare}. (A special case arises from considering
integers, which corresponds to choosing grid locations.)
Our approximation works online, i.e., we choose the cities
in a specified order, and no changes can be made to previously allocated
cities; clearly, this implies approximation factors for the corresponding offline
problems.

There are lower bounds for $\max_{1 \le i \le k}\phi(C_i)$ that generally cannot
be achieved by any algorithm. One important result is the following theorem.

\begin{theo}\label{theo-lower-bound-cont}
Let $C$ be any city. Then $\phi(C) \ge 0.650245$.
\end{theo}

A proof can be found in \cite{bender}. For $n_1=1$ any algorithm must select
the whole unit square, thus $2/3$, the $\phi$-value of a square, is a lower
bound for the achievable $\phi$-value. We will discuss better lower bounds in the conclusions.

\section{An Allocation Strategy}
While long and narrow shapes tend to have large $\phi$-values, shapes that fill
large parts of an enclosing rectangle with similar width and height usually
have better average distances; however, one has to make sure that early choices
with small average distance do not 
leave narrow pieces with high average distance, or even disconnected pieces,
making the normalized $\phi$-values potentially unbounded. 

\begin{figure}
  \centerline{\includegraphics[width=.28\textwidth]{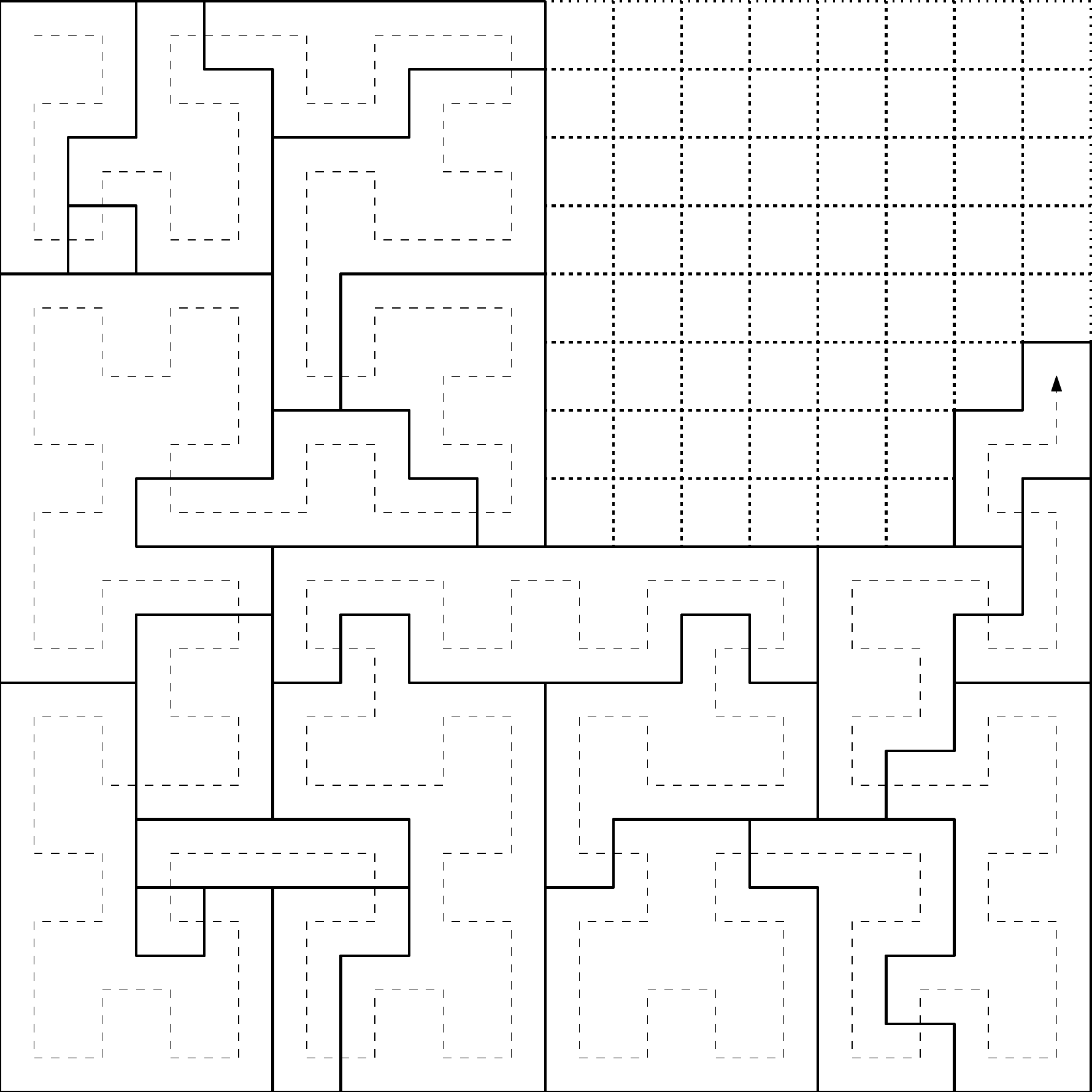}}
  \caption{A sample allocation according to our strategy.}\label{fig-allocation}
\end{figure}

Our approach uses the recursive Hilbert family of curves
in order to yield a provably constant competitive factor.
That family is based on a recursive construction scheme
and becomes space-filling for infinite repetition of said scheme \cite{sagan}.
For a finite number $r$ of repetitions, the curve traverses all points of the used grid. For
$1 \le r \le 3$, the curve is shown in Fig.~\ref{fig-hilbert-curve}. Thus, the
Hilbert curve provides an order for the cells of the grid, which is then used for
allocation, as illustrated in Fig.~\ref{fig-allocation}.
More formal details of the recursive definition of the Hilbert family
(e.g.\ with text-rewriting rules, such
as the ones in \cite{hilb-grammar})
are cited and sketched in the following Section~\ref{sec:tech}. 

\setlength{\unitlength}{0.2cm}

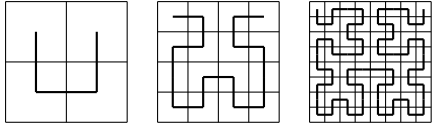
\begin{figure}[htp]
\centering
\begin{picture}(28,8)

\thinlines
\multiput(0,0)(4,0){3}{\line(0,1){8}}
\multiput(0,0)(0,4){3}{\line(1,0){8}}

\thicklines
\put(2,2){\line(0,1){4}}
\put(2,2){\line(1,0){4}}
\put(6,2){\line(0,1){4}}

\thinlines
\multiput(10,0)(2,0){5}{\line(0,1){8}}
\multiput(10,0)(0,2){5}{\line(1,0){8}}

\thicklines
\put(11,7){\line(1,0){2}}
\put(13,5){\line(0,1){2}}
\put(11,5){\line(1,0){2}}
\put(11,3){\line(0,1){2}}
\put(11,1){\line(0,1){2}}
\put(11,1){\line(1,0){2}}
\put(13,1){\line(0,1){2}}
\put(13,3){\line(1,0){2}}
\put(15,1){\line(0,1){2}}
\put(15,1){\line(1,0){2}}
\put(17,1){\line(0,1){2}}
\put(17,3){\line(0,1){2}}
\put(15,5){\line(1,0){2}}
\put(15,5){\line(0,1){2}}
\put(15,7){\line(1,0){2}}

\thinlines
\multiput(20,0)(1,0){9}{\line(0,1){8}}
\multiput(20,0)(0,1){9}{\line(1,0){8}}

\thicklines
\put(20.5,6.5){\line(0,1){1}}
\put(20.5,6.5){\line(1,0){1}}
\put(21.5,6.5){\line(0,1){1}}
\put(21.5,7.5){\line(1,0){1}}
\put(22.5,7.5){\line(1,0){1}}
\put(23.5,6.5){\line(0,1){1}}
\put(22.5,6.5){\line(1,0){1}}
\put(22.5,5.5){\line(0,1){1}}
\put(22.5,5.5){\line(1,0){1}}
\put(23.5,4.5){\line(0,1){1}}
\put(22.5,4.5){\line(1,0){1}}
\put(21.5,4.5){\line(1,0){1}}
\put(21.5,4.5){\line(0,1){1}}
\put(20.5,5.5){\line(1,0){1}}
\put(20.5,4.5){\line(0,1){1}}
\put(20.5,3.5){\line(0,1){1}}
\put(20.5,3.5){\line(1,0){1}}
\put(21.5,2.5){\line(0,1){1}}
\put(20.5,2.5){\line(1,0){1}}
\put(20.5,1.5){\line(0,1){1}}
\put(20.5,0.5){\line(0,1){1}}
\put(20.5,0.5){\line(1,0){1}}
\put(21.5,0.5){\line(0,1){1}}
\put(21.5,1.5){\line(1,0){1}}
\put(22.5,0.5){\line(0,1){1}}
\put(22.5,0.5){\line(1,0){1}}
\put(23.5,0.5){\line(0,1){1}}
\put(23.5,1.5){\line(0,1){1}}
\put(22.5,2.5){\line(1,0){1}}
\put(22.5,2.5){\line(0,1){1}}
\put(22.5,3.5){\line(1,0){1}}
\put(23.5,3.5){\line(1,0){1}}
\put(24.5,3.5){\line(1,0){1}}
\put(25.5,2.5){\line(0,1){1}}
\put(24.5,2.5){\line(1,0){1}}
\put(24.5,1.5){\line(0,1){1}}
\put(24.5,0.5){\line(0,1){1}}
\put(24.5,0.5){\line(1,0){1}}
\put(25.5,0.5){\line(0,1){1}}
\put(25.5,1.5){\line(1,0){1}}
\put(26.5,0.5){\line(0,1){1}}
\put(26.5,0.5){\line(1,0){1}}
\put(27.5,0.5){\line(0,1){1}}
\put(27.5,1.5){\line(0,1){1}}
\put(26.5,2.5){\line(1,0){1}}
\put(26.5,2.5){\line(0,1){1}}
\put(26.5,3.5){\line(1,0){1}}
\put(27.5,3.5){\line(0,1){1}}
\put(27.5,4.5){\line(0,1){1}}
\put(26.5,5.5){\line(1,0){1}}
\put(26.5,4.5){\line(0,1){1}}
\put(25.5,4.5){\line(1,0){1}}
\put(24.5,4.5){\line(1,0){1}}
\put(24.5,4.5){\line(0,1){1}}
\put(24.5,5.5){\line(1,0){1}}
\put(25.5,5.5){\line(0,1){1}}
\put(24.5,6.5){\line(1,0){1}}
\put(24.5,6.5){\line(0,1){1}}
\put(24.5,7.5){\line(1,0){1}}
\put(25.5,7.5){\line(1,0){1}}
\put(26.5,6.5){\line(0,1){1}}
\put(26.5,6.5){\line(1,0){1}}
\put(27.5,6.5){\line(0,1){1}}

\end{picture}
\caption{Hilbert curve with $1 \le r \le 3$.}\label{fig-hilbert-curve}
\end{figure}

More technically, the unit square is recursively subdivided into a grid consisting 
of $2^r\times 2^r$ grid cells, for an appropriate refinement level $r>0$, 
as shown in Fig.~\ref{fig-hilbert-curve}.
For the sake of
concise presentation, we assume that every input 
$n_i$ is an integral multiple of $c=4^{-R}$, for an appropriately large $R>0$.
(We will
mention in the Conclusions how this assumption can be removed, based
on Lemma~\ref{lemma-wi-bound2}.)
Similar to the recursive structure of 
quad-trees, the actual subdivision can be performed in 
a self-refining manner, whenever a grid cell is
not completely filled. 
This means that during the course of the online allocation,
we may use different refinement levels in different parts of the layout;
however, this will not affect the overall analysis, as further refinement of the grid
does not change the quality of existing shapes. 

\begin{defi}
For a give refinement level $r$, an $r$-{\em pixel} $P$ is a grid square of size $2^{-r}\times 2^{-r}$.
For a given allocated shape $C_i$, a pixel is {\em full} if $P\subseteq C_i$;
it is {\em fractional}, if $P\cap C_i\neq\emptyset$ and $P\not\subset C_i$.
\end{defi}

Now the description of the algorithm is simple: 
for every input $n_i$, choose the next set of $n_i/2^R$ $R$-pixels traversed by the Hilbert curve
as the city $C_i$, starting in the upper left corner of the grid. For an illustration,
see Fig.~\ref{fig-allocation}.

The following lemma is a consequence of the recursive structure of the Hilbert family; see
the following Section~\ref{sec:tech} for a formal argument.
We use it in Section~\ref{sec:ana} for deriving upper bounds.

\begin{lemm}\label{lemma-wi-bound}
Let $C$ be a city generated by our strategy with area at most $n \le l \, 4^j \, 4^{-R}$
for $j \in \{0,1,\ldots,R\}, l \in \mathbb{N}$.
Then at any refinement level $r$, $C$ contains at most two fractional $r$-pixels.
\end{lemm}

\section{Technical Details of Shape Allocation}
\label{sec:tech}

A technical different description of the Hilbert family can be
based on the string representation given in \cite{hilb-grammar}.
There, the authors use the following recursion, based on the letters u, r, d, l for denoting
``up'', ``right'', ``down'', and ``left''.

\begin{eqnarray*}
y_1 & = & \mathrm{urd}\\
y_n & = & h_4(y_{n-1})\mathrm{u}y_{n-1}\mathrm{r}y_{n-1}\mathrm{d}h_5(y_{n-1}) \mathrm{\mbox{ if }} n>1
\end{eqnarray*}%

where $h_4$ and $h_5$ are defined as%

\begin{equation*}
h_4(x) =
\begin{cases}
\text{r} & \text{if } x=\text{u},\\
\text{l} & \text{if } x=\text{d},\\
\text{u} & \text{if } x=\text{r},\\
\text{d} & \text{if } x=\text{l}.
\end{cases} \mathrm{\mbox{ and }}
h_5(x) = 
\begin{cases}
\text{l} & \text{if } x=\text{u},\\
\text{r} & \text{if } x=\text{d},\\
\text{d} & \text{if } x=\text{r},\\
\text{u} & \text{if } x=\text{l}.
\end{cases}
\end{equation*}%

and $h_i(x_0x_1x_2\ldots)=h_i(x_0)h_i(x_1)h_i(x_2)\ldots$, $i\in \{4,5\}$.

We combine those characters to four sequences of length three and build an
L-system, a special kind of formal grammar with parallel rewriting.
Let $A=\mathrm{urd}$, $B=\mathrm{ldr}$, $C=\mathrm{rul}$, and $D=\mathrm{dlu}$, as shown in Figure~\ref{fig-abcd}
Using the production rules
\begin{eqnarray*}
A & \rightarrow & CAAB,\\
B & \rightarrow & DBBA,\\
C & \rightarrow & ACCD\mathrm{\mbox{ and }}\\
D & \rightarrow & BDDC
\end{eqnarray*}
and starting with $A$, you get the same string from \cite{hilb-grammar}
if you replace $X_0X_1X_2X_3$ by $X_0uX_1rX_2dX_3$, $X_0,X_1,X_2,X_3 \in \{A,B,C,D\}$, after each
use of a production rule.

A sequence of symbols produced by the L-system can be interpreted graphically
as a sequence of sub-squares $E_{r-1}^2[j][k], j,k \in \{1,\ldots,2^{r-1}\}$,
see Figure~\ref{fig-abcd}. The figure also shows that $A$ is $D$ rotated
by $180^{\circ}$. The same is true for $B$ and $C$.
We denote by $\overline{X}$ the symbol that is $X$ rotated by $180^{\circ}$.
As we are only interested in the shape of cities, we identify symmetric cities.
Consequently, we make no distinction
between $X$ and $\overline{X}$ when we look at a single symbol. We write
$X \equiv \overline{X}$. Similarly, we get equivalences for longer
sequences of symbols: $XY \equiv YX$, because the order of the successive
sub-squares $E_{r-1}^2[j][k]$ does not change the shape. Furthermore,
we have $X\overline{X} \equiv Y\overline{Y}$ for each $X$ and $Y$, as it is
always the simple shape followed by the rotated shape, and $XX \equiv \overline{X}\overline{X}$
for two occurrences of the same shape in succession.

\setlength{\unitlength}{1cm}

\newsavebox{\squareIconB}
\savebox{\squareIconB}(1,1)[bl]{
\put(0,0){\line(1,0){1}}
\put(0,0){\line(0,1){1}}
\put(0,1){\line(1,0){1}}
\put(1,0){\line(0,1){1}}
}

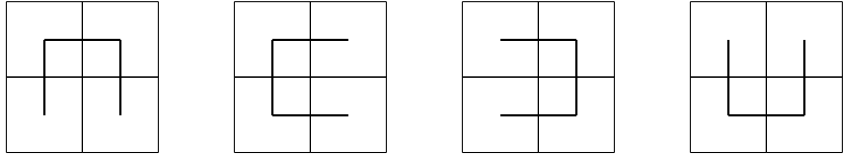
\begin{figure}[bp]
\centering
\begin{picture}(11,2)

\thinlines
\put(0,0){\usebox{\squareIconB}}
\put(0,1){\usebox{\squareIconB}}
\put(1,0){\usebox{\squareIconB}}
\put(1,1){\usebox{\squareIconB}}

\thicklines
\put(0.5,1.5){\line(0,-1){1}} 
\put(0.5,1.5){\line(1,0){1}}  
\put(1.5,1.5){\line(0,-1){1}} 

\thinlines
\put(3,0){\usebox{\squareIconB}}
\put(3,1){\usebox{\squareIconB}}
\put(4,0){\usebox{\squareIconB}}
\put(4,1){\usebox{\squareIconB}}

\thicklines
\put(3.5,1.5){\line(0,-1){1}} 
\put(3.5,0.5){\line(1,0){1}}  
\put(3.5,1.5){\line(1,0){1}}  

\thinlines
\put(6,0){\usebox{\squareIconB}}
\put(6,1){\usebox{\squareIconB}}
\put(7,0){\usebox{\squareIconB}}
\put(7,1){\usebox{\squareIconB}}

\thicklines
\put(6.5,0.5){\line(1,0){1}}  
\put(6.5,1.5){\line(1,0){1}}  
\put(7.5,1.5){\line(0,-1){1}} 

\thinlines
\put(9,0){\usebox{\squareIconB}}
\put(9,1){\usebox{\squareIconB}}
\put(10,0){\usebox{\squareIconB}}
\put(10,1){\usebox{\squareIconB}}

\thicklines
\put(9.5,1.5){\line(0,-1){1}}  
\put(9.5,0.5){\line(1,0){1}}   
\put(10.5,1.5){\line(0,-1){1}} 

\end{picture}
\caption{$A$, $B$, $C$, and  $D$}\label{fig-abcd}
\end{figure}

\begin{lemm}\label{lemma-replace}
After $r>1$ uses of the production rules (where $r=1$ means that we are still at the start symbol $A$),
the resulting sequence contains all sub-sequences of length $2^{r-2}$ that can be created with
the L-system, except for symmetry.
\end{lemm}

{\bf Proof:} 
We prove the claim via induction over $r$: \mbox{}
\begin{description}
\item[$r=2:$ ] $CAAB$ contains $C$, $B$, and $A \equiv \overline{A}=D$, i.e.,
all sequences of length 1.
\item[$r=3:$ ] $ACCD CAAB CAAB DBBA$ contains
\begin{itemize}
\item $AA \equiv DD$
\item $AB \equiv BA \equiv DC \equiv CD$
\item $AC \equiv CA \equiv DB \equiv BD$
\item $BC \equiv CB \equiv AD \equiv DA$
\item $BB \equiv CC$,
\end{itemize}
i.e, all possible sequences of length 2.

\item[$r \rightarrow r+1, r \ge 3:$ ] A sequence of length $2^{r-1}$ has been created
from a sequence of length at most $2^{r-1}/4+1=2^{r-3}+1$, because every symbol is replaced
by exactly 4 symbols. If the claim holds for $r \ge 3$, then all sequences of length
$2^{r-2}$ have been created, i.e., all sequences of length $2^{r-3}+1$ have been created as
well. Thus, in the next step the production rules are applied to all possible sequences of
that length, yielding all sequences of length $2^{r-1}$ that can be created with the L-system.\hfill\qed
\end{description}
\hfill $\Box$

\begin{lemm}\label{lemma-number-4s}
A city $W_n$ contains at most parts of $\lceil n/4 \rceil + 1$ sub-squares 
$E_{r-1}^2[j][k]$, $j,k \in \{1,\ldots,2^{r-1}\}$, i.e., sub-squares that consist
of exactly 4 cells.
\end{lemm}

{\bf Proof: } 
A city $W_n$ consists of exactly $n$ cells. Assume that these cells belong to at least
$\lceil n/4 \rceil + 2$ sub-squares $E_{r-1}^2[j][k]$. Then at least three of those
sub-squares cannot belong to $W_n$ as a whole but only in part, because $n$ cells cannot
completely fill more than $\lceil n/4 \rceil$ of the sub-squares, and if two more are partially
filled not even all those can be filled completely. Consider the sequence of sub-squares of $W_n$ in
the order given by the Hilbert curve. One of the sub-squares that is not the first or the last in
the sequence cannot completely belong to $W_n$. This is a contradiction to the definition
of the Hilbert curve, which recursively uses the same construction scheme for sub-squares
on every level of refinement. \hfill\qed

\begin{lemm}\label{lemma-sequences}
When the L-system has generated all sequences of length $\lceil n/4 \rceil + 1$, the
resulting Hilbert curve traverses a city $W_n$.
\end{lemm}

{\bf Proof: } 
Each symbol of the L-system corresponds to a sub-square $E_{r-1}^2[j][k]$. Once all
sequences of length $\lceil n/4 \rceil + 1$ symbols have been generated, all possible
cities consisting of $\lceil n/4 \rceil + 1$ sub-squares have been generated, too. With
Lemma~\ref{lemma-number-4s} the claim follows.\hfill \qed

\begin{lemm}
For $r = \lceil \log_2(\lceil n/4 \rceil + 1)\rceil + 2$ the Hilbert curve
traverses a city $W_n$.
\end{lemm}

{\bf Proof: } 
Using Lemmas~\ref{lemma-replace} and \ref{lemma-sequences} we know that the Hilbert curve
traverses a city $W_n$, if the following holds:

\[
2^{r-2} \ge \lceil n/4 \rceil + 1
\]

This is true for

\[
r = \lceil \log_2(\lceil n/4 \rceil + 1)\rceil + 2.
\]\hfill \qed


\section{Analysis}
\label{sec:ana}
For the analysis of our allocation scheme we will first make use of Lemma~\ref{lemma-wi-bound}. 
As noted in Lemma~\ref{lemma-wi-bound2}, filling in the two fractional pixels
of an allocated shape yields an estimate for the total distance at a coarser
refinement level. In a second step, this will be used to derive global
bounds by computing the worst-case bounds for shapes of at most refinement level 3;
thus the presented computational results for shapes of limited size are not
merely experiments, but yield a general upper bound on the competitive factor. 
(As discussed in the Conclusions, carrying out the computations on a lower or
higher refinement level gives looser or tighter results.)

\bigskip
In the following, we denote by $W_n$ the worst city consisting 
of $n$ pixels that can be produced by our Hilbert strategy; because of the normalized
nature of $\phi$, this is independent on the size of the pixels, and only the shape matters.

\begin{lemm}\label{lemma-wi-bound2}
Let $C$ be a city generated by our strategy with area at most $n \le l \, 4^r \, 4^{-R}$
for $r \in \{0,1,\ldots,R\}, l \in \mathbb{N}$.
Then we have $c(C) \le c(W_{l+1})$, where 
$W_{l+1}$ is a worst case among all cities produced by our allocation scheme 
that consists of $(l+1)$ $r$-pixels.
\end{lemm}

{\bf Proof: } 
By Lemma~\ref{lemma-wi-bound}, we know that only the first and the last
pixel of $C$ may be fractional. Therefore
$C$ cannot intersect more than $l+1$ $r$-pixels. By replacing the two fractional
pixels by full pixels, we get a city $W$ that consists of $l+1$ full $r$-pixels,
and $c(C) \le c(W)$. By definition, $c(W) \le c(W_{l+1})$, and the claim holds.

\begin{figure}[h!]
  \centerline{\includegraphics[width=.5\textwidth]{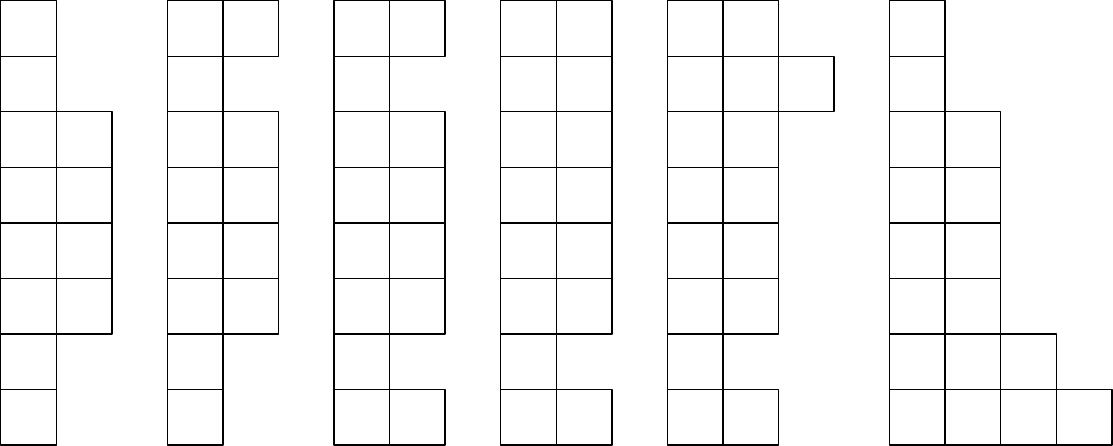}}
  \caption{Worst cases $W_n$ for $12 \le n \le 17$.}\label{fig-wi}
\end{figure}

Therefore, we can give upper bounds for the worst case by considering the
values of $W_n$ at some moderate refinement level. 
The $W_n$ can be found by enumeration; as described by the technical lemmas in preceding section,
a speed-up can be achieved by
making use of the recursive construction of the $W_n$. We
determined the shapes and $\phi$-values of the $W_n$ for $n \le 65$;
by Lemma~\ref{lemma-wi-bound2}, this suffices to provide upper bounds for
all cities with area up to $64*2^{-r}$, i.e., these computational results
give an estimate for the round-up error using refinement level 3.
The full table of average distances for this refinement level can be seen in Table~\ref{tbl-phi-wi};
the worst cases among the examined ones are $W_{56}$ and $W_{14}$, which have the
same shape, shown in Fig.~\ref{fig-wi}.

\begin{table}[h!tb]
\centering
\scriptsize
\caption{Total and average distances for cities $W_n$ allocated according to our strategy, as well as
the optimal values $c_{city}(n)$ according to \cite{dfr-psmd-11}.}\label{tbl-phi-wi}
\begin{tabular}[c]{|r|r|r|r|r||r|r|r|r|r|}
\hline
$n$ & $c^*(W_n)$ & $c_{{city}}(n)$ & $\phi(W_n)$ & $\Phi(W_{n})$ & $n$ & $c^*(W_n)$ & $c_{{city}}(n)$ & $\phi(W_n)$ & $\Phi(W_{n})$\\
\hline \hline
$1$ & $0 \, ^1/_3$ & $0 \, ^1/_3$ & $0.6667$ & & $34$ & $2835 \, ^2/_3$ & $2216 \, ^2/_3$ & $0.8414$ & $0.9691$\\
\hline
$2$ & $2$ & $2$ & $0.7071$ & & $35$ & $3045$ & $2384$ & $0.8403$ & $0.9712$\\
\hline
$3$ & $6$ & $5 \, ^2/_3$ & $0.7698$ & & $36$ & $3266$ & $2554 \, ^2/_3$ & $0.8400$ & $0.9772$\\
\hline
$4$ & $13 \, ^1/_3$ & $10 \, ^2/_3$ & $0.8333$ & & $37$ & $3519 \, ^1/_3$ & $2727 \, ^2/_3$ & $0.8453$ & $0.9726$\\
\hline
$5$ & $24$ & $19 \, ^2/_3$ & $0.8587$ & & $38$ & $3799 \, ^1/_3$ & $2921 \, ^2/_3$ & $0.8536$ & $0.9682$\\
\hline
$6$ & $38$ & $30$ & $0.8619$ & & $39$ & $4049 \, ^2/_3$ & $3117 \, ^2/_3$ & $0.8527$ & $0.9570$\\
\hline
$7$ & $50 \, ^2/_3$ & $44$ & $0.7816$ & & $40$ & $4309 \, ^1/_3$ & $3322$ & $0.8517$ & $0.9463$\\
\hline
$8$ & $74 \, ^2/_3$ & $61 \, ^1/_3$ & $0.8250$ & & $41$ & $4545$ & $3530 \, ^1/_3$ & $0.8445$ & $0.9307$\\
\hline
$9$ & $102$ & $81$ & $0.8395$ & & $42$ & $4788$ & $3749$ & $0.8376$ & $0.9214$\\
\hline
$10$ & $134 \, ^2/_3$ & $106 \, ^1/_3$ & $0.8517$ & & $43$ & $5009$ & $3976$ & $0.8262$ & $0.9306$\\
\hline
$11$ & $162$ & $135 \, ^2/_3$ & $0.8074$ & & $44$ & $5266 \, ^2/_3$ & $4205 \, ^1/_3$ & $0.8202$ & $0.9393$\\
\hline
$12$ & $210 \, ^2/_3$ & $165 \, ^1/_3$ & $0.8446$ & & $45$ & $5641 \, ^1/_3$ & $4456 \, ^2/_3$ & $0.8306$ & $0.9393$\\
\hline
$13$ & $262 \, ^2/_3$ & $203$ & $0.8621$ & & $46$ & $6031 \, ^1/_3$ & $4712$ & $0.8405$ & $0.9395$\\
\hline
$14$ & $322$ & $244$ & $0.8781$ & & $47$ & $6379 \, ^2/_3$ & $4970 \, ^1/_3$ & $0.8425$ & $0.9439$\\
\hline
$15$ & $371 \, ^2/_3$ & $290 \, ^2/_3$ & $0.8530$ & & $48$ & $6741 \, ^1/_3$ & $5234$ & $0.8446$ & $0.9505$\\
\hline
$16$ & $434 \, ^2/_3$ & $338 \, ^2/_3$ & $0.8490$ & $1.1764$ & $49$ & $7147 \, ^1/_3$ & $5507 \, ^1/_3$ & $0.8505$ & $0.9512$\\
\hline
$17$ & $512 \, ^2/_3$ & $396 \, ^1/_3$ & $0.8605$ & $1.1497$ & $50$ & $7586 \, ^1/_3$ & $5788$ & $0.8583$ & $0.9516$\\
\hline
$18$ & $602 \, ^1/_3$ & $457 \, ^1/_3$ & $0.8764$ & $1.1174$ & $51$ & $7993$ & $6076 \, ^1/_3$ & $0.8606$ & $0.9559$\\
\hline
$19$ & $685$ & $522 \, ^1/_3$ & $0.8706$ & $1.1058$ & $52$ & $8411 \, ^1/_3$ & $6368$ & $0.8628$ & $0.9620$\\
\hline
$20$ & $768$ & $591 \, ^2/_3$ & $0.8587$ & $1.1098$ & $53$ & $8878$ & $6691 \, ^2/_3$ & $0.8683$ & $0.9619$\\
\hline
$21$ & $870$ & $663$ & $0.8610$ & $1.0903$ & $54$ & $9379 \, ^1/_3$ & $7017 \, ^1/_3$ & $0.8754$ & $0.9617$\\
\hline
$22$ & $992 \, ^2/_3$ & $749 \, ^1/_3$ & $0.8745$ & $1.0713$ & $55$ & $9835$ & $7352 \, ^1/_3$ & $0.8768$ & $0.9569$\\
\hline
$23$ & $1101 \, ^2/_3$ & $839 \, ^1/_3$ & $0.8685$ & $1.0424$ & $56$ & $10304$ & $7690$ & $0.8781$ & $0.9522$\\
\hline
$24$ & $1216$ & $933$ & $0.8619$ & $1.0150$ & $57$ & $10733$ & $8033 \, ^2/_3$ & $0.8751$ & $0.9445$\\
\hline
$25$ & $1322 \, ^1/_3$ & $1032\, ^1/_3$ & $0.8463$ & $0.9777$ & $58$ & $11173 \, ^1/_3$ & $8384 \, ^2/_3$ & $0.8723$ & $0.9372$\\
\hline
$26$ & $1432$ & $1134$ & $0.8309$ & $0.9701$ & $59$ & $11583 \, ^2/_3$ & $8749 \, ^1/_3$ & $0.8665$ & $0.9268$\\
\hline
$27$ & $1527 \, ^2/_3$ & $1249$ & $0.8066$ & $0.9785$ & $60$ & $12005 \, ^1/_3$ & $9117 \, ^1/_3$ & $0.8610$ & $0.9225$\\
\hline
$28$ & $1672$ & $1365 \, ^1/_3$ & $0.8061$ & $0.9864$ & $61$ & $12391$ & $9506 \, ^1/_3$ & $0.8527$ & $0.9232$\\
\hline
$29$ & $1853 \, ^1/_3$ & $1492$ & $0.8184$ & $0.9773$ & $62$ & $12862$ & $9904 \, ^2/_3$ & $0.8499$ & $0.9201$\\
\hline
$30$ & $2046$ & $1622 \, ^2/_3$ & $0.8301$ & $0.9710$ & $63$ & $13415$ & $10305 \, ^1/_3$ & $0.8517$ & $0.9175$\\
\hline
$31$ & $2213$ & $1759 \, ^2/_3$ & $0.8272$ & $0.9726$ & $64$ & $13924 \, ^2/_3$ & $10719 \, ^1/_3$ & $0.8499$ & \\
\hline
$32$ & $2393 \, ^1/_3$ & $1898 \, ^2/_3$ & $0.8263$ & $0.9791$ & $65$ & $14452 \, ^2/_3$ & $11139 \, ^2/_3$ & $0.8486$ & \\
\hline
$33$ & $2602$ & $2057$ & $0.8319$ & $0.9735$ & & & & &\\
\hline
\end{tabular}
\end{table}

%




\begin{theo}\label{theo-pho-bound}
Our strategy guarantees $\max_{1 \le n \le k}{\phi(C_n)} \le 1.1764$.
\end{theo}






{\bf Proof: } 
Consider a city $C$ of size $n$ generated by our strategy. If $n$ is sufficiently small, i.e.,
smaller than an $R-r$-pixel, $r \ge 0$, $C$ consists of at most $4^{r}$ cells and its average
distance can be bounded by the worst case for that
particular number of cells. In the case that $C$ has a larger, more complicated shape, an analysis of
a finite number of shapes is still sufficient:

We know that $n > 4^{j}c$ and can assume that
$n \le 4^{j+1}c$ (or else we use the analysis on the less refined $E_{r-(j+1)}^2[p][q]$). Thus, there must
be an $l$ such that $l4^{j}c < n \le (l+1)4^{j}c$ with $l = 1,\ldots,3$. Yet, we can get closer to
$n$, as we know that $E_{r-j}^2[p][q]$ consists of $4^{j}$ cells. We get the inequality $l 4^{j-k} < n \le (l+1)4^{j-k}c$,
$k \le j$, $l = 4^k,\ldots,4^{k+1}-1$.

Hence, a city of arbitrary size $n$ corresponds to at most $(l+1)$ sub-squares of a certain size (depending
on the precision of the analysis), i.e., a city of size at most $(l+1)4^{j-k}c$. Now we can use Lemma~\ref{lemma-wi-bound} to bound the average
distance of the city, yielding

\begin{eqnarray}
\phi(C) \le \frac{2 \, c(W)}{(l \, 4^{j-k}c)^{5/2}} & = & \frac{\phi(W_{l+2})((l+2)
\, 4^{j-k}c)^{5/2}}{(l \, 4^{j-k}c)^{5/2}}\\
& = &
\phi(W_{l+2})\left(1+\frac{2}{l}\right)^{5/2} =: \Phi(W_l).  
\end{eqnarray}
The resulting bound is $\max(\{\phi(W_i): 1\le i\le 4^j\} \cup \{\Phi(W_l): 4^k \le l \le 4^{k+1}-1\})$. Note that the number of shapes
considered is at most $4^{k+1}$.

We conducted the calculations for $k=2$ and list the results in Table~\ref{tbl-phi-wi}. As it turns out, the maximum is
attained for $\Phi(W_{16})=1.1764$. \hfill \qed

\begin{coro}
Our strategy achieves a competitive factor of 1.8092.
\end{coro}

{\bf Proof: } 
According to Theorem~\ref{theo-lower-bound-cont}, no algorithm can guarantee
a better $\phi$-value than 0.650245. Our strategy yields an upper bound
of 1.1764. This results in a factor of $1.1764/0.650245 \approx 1.8092$.
\hfill\qed

\section{Discrete Point Sets}
Our above analysis relies on continuous weight distributions, which
imply the lower bound on $\phi$-values stated in Theorem~1.
This does not include the discrete scenario, in which the values $n_i$ indicate
a number of integer grid points that have to be chosen from an appropriate $N\times N$-grid. 
As discussed in the paper~\cite{dfr-psmd-11}, considering discrete
weight distributions may allow lower average distances; e.g.,
a single point yields a $\phi$-value of 0. As a consequence,
{\em towns} (subsets of the integer grid) have lower average
distances than cities of the same total weight. However,
we still get a competitive ratio for the case of online
towns.

\begin{theo}\label{theo-city-comp}
For $n$-towns, a Hilbert-curve strategy guarantees a competitive factor of at most 1.7848 for the $\phi$-value.
\end{theo}

{\bf Proof: } 
Lemma~\ref{lemma-wi-bound} still holds, so analogously to Theorem~\ref{theo-pho-bound}, 
we consider the values up to $n=64$, and 
show that the worst case is attained for $n=16$, which yields an upper bound of
1.123. See Table~\ref{tbl-towns} for detailed numbers.

For a lower bound, the general value of 0.650245 for $\phi$-values cannot be applied,
as discrete point sets may have lower average distance. Instead, we verify that the ratio $\rho(n)$ of
achieved $\phi$ to optimal $\phi$, is less than 1.7848; 
this is the same as $c(T_n)/c_{{town}}(n)$ for $n\leq 64 $,
see Table~\ref{tbl-towns}.
For $65\leq n\leq 80$, the optimal values in~\cite{dfr-psmd-11}
allow us to verify that $\phi\geq 0.6292$; see our Table~\ref{tbl-opt-phi}.

\begin{figure}[h]
  \begin{center}
  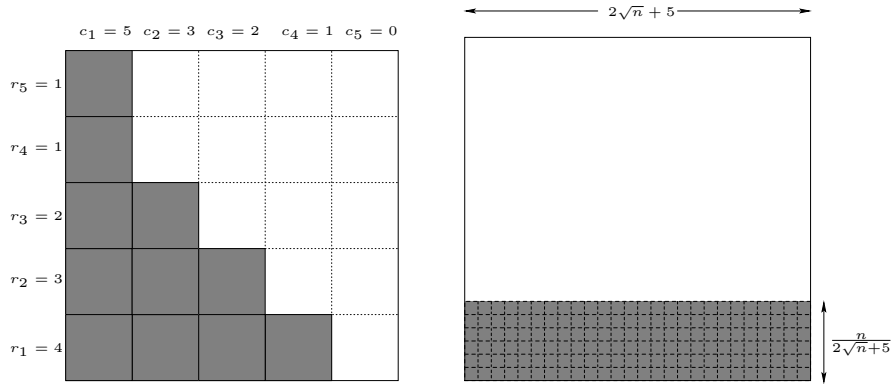
  \caption{Establishing a lower bound for $\phi$: Defining $\Lambda(n)$; an arrangement that maximizes $\Lambda(n)$.}\label{fig-corr}
  \end{center}
\end{figure}

Thus, we have to establish
a lower bound for $\phi$ for $n\geq 81$. We 
make use of equation (5), p.~89 of \cite{dfr-psmd-11}; see Fig.~\ref{fig-corr}:
for a given number $n$ of grid points, the difference between the optimal total
Manhattan distance $c_{{city}}(n)$ 
for a city consisting of $n$ unit squares
and the optimal total distance
$c_{{town}}(n)$ for a town consisting of $n$ grid points is equal to
$\Lambda(n):=\frac{1}{6}\left(\sum_i c_i^2 + \sum_j r_j^2\right),$
where $c_i$ is the number of grid points in column $i$, and $r_j$ is the number of 
grid points in row $j$. Because $\frac{2c_{{city}}(n)}{n^{2.5}}$ is bounded from
below by $\psi=0.650245$, we get a lower bound of
$\psi-\frac{2\Lambda(n)}{n^{2.5}}\leq \frac{2c_{{town}}(n)}{n^{2.5}}$
for the $\phi$-value of an $n$-town.

This leaves the task of providing an upper bound for $2\Lambda(n)/n^{2.5}$.
According to Lemma~5 of \cite{dfr-psmd-11}, the bounding box of an optimal $n$-town does
not exceed $2\sqrt{n}+5$. Therefore, we have $c_i\leq 2\sqrt{n}+5$; as $\sum_i c_i =n$ and
the function $\sum_i c_i^2$ is superlinear in the $c_i$, we conclude that $\sum_i c_i^2$
is maximized by subdividing $n$ into $\frac{n}{2\sqrt{n}+5}$ columns of $2\sqrt{n}+5$
points each, so $\sum_i c_i^2\leq n(2\sqrt{n}+5)$. Analogously, we have
$\sum_j r_j^2\leq n(2\sqrt{n}+5)$, so $2\Lambda(n)/n^{2.5}\leq \frac{2}{3} (\frac{2}{n}+\frac{5}{n^{1.5}})$.
For $n\geq 81$, this implies $2\Lambda(n)/n^{2.5}\leq \frac{4}{243}+\frac{10}{2187}=0.0210333...$
or $\phi(n)\geq 0.6292$.
This yields an overall competitive ratio of not more than 1.123/0.6292, i.e., 1.7848.
\hfill\qed

\begin{table}[h!tb]
\centering
\scriptsize
\caption{Total distances for towns $T_n$ allocated according to our strategy, and 
the corresponding optimal values $c_{town}$ for $n$-towns.}
\label{tbl-towns}
\begin{tabular}[c]{|r|r|r|r|r|r||r|r|r|r|r|r|}
\hline
$n$ & $c(T_n)$ & $c_{town}(n)$ & $\rho(n)$ & $\phi(T_n)$ & $\Phi(T_{n})$ & $n$ & $c(T_n)$ & $c_{town}(n)$ & $\rho(n)$ & $\phi(T_n)$ & $\Phi(T_{n})$\\
\hline \hline
$1$ & $0$ & $0$ & $-$ & $0.0000$ & & $34$ & $2763$ & $2153$ & $1.2833$ & $0.8198$ & $0.9453$\\
\hline
$2$ & $1$ & $1$ & $1.0000$ & $0.3536$ & & $35$ & $2968$ & $2318$ & $1.2804$ & $0.8191$ & $0.9482$\\
\hline
$3$ & $4$ & $4$ & $1.0000$ & $0.5132$ & & $36$ & $3186$ & $2486$ & $1.2816$ & $0.8194$ & $0.9550$\\
\hline
$4$ & $10$ & $8$ & $1.2500$ & $0.6250$ & & $37$ & $3436$ & $2656$ & $1.2937$ & $0.8252$ & $0.9511$\\
\hline
$5$ & $20$ & $16$ & $1.2500$ & $0.7155$ & & $38$ & $3713$ & $2847$ & $1.3042$ & $0.8342$ & $0.9473$\\
\hline
$6$ & $33$ & $25$ & $1.3200$ & $0.7485$ & & $39$ & $3960$ & $3040$ & $1.3026$ & $0.8338$ & $0.9366$\\
\hline
$7$ & $44$ & $38$ & $1.1579$ & $0.6788$ & & $40$ & $4216$ & $3241$ & $1.3008$ & $0.8333$ & $0.9263$\\
\hline
$8$ & $66$ & $54$ & $1.2222$ & $0.7292$ & & $41$ & $4448$ & $3446$ & $1.2908$ & $0.8265$ & $0.9112$\\
\hline
$9$ & $92$ & $72$ & $1.2778$ & $0.7572$ & & $42$ & $4687$ & $3662$ & $1.2799$ & $0.8200$ & $0.9017$\\
\hline
$10$ & $123$ & $96$ & $1.2812$ & $0.7779$ & & $43$ & $4904$ & $3886$ & $1.2620$ & $0.8089$ & $0.9112$\\
\hline
$11$ & $148$ & $124$ & $1.1935$ & $0.7376$ & & $44$ & $5154$ & $4112$ & $1.2534$ & $0.8027$ & $0.9203$\\
\hline
$12$ & $194$ & $152$ & $1.2763$ & $0.7778$ & & $45$ & $5524$ & $4360$ & $1.2670$ & $0.8133$ & $0.9205$\\
\hline
$13$ & $244$ & $188$ & $1.2979$ & $0.8009$ & & $46$ & $5909$ & $4612$ & $1.2812$ & $0.8235$ & $0.9212$\\
\hline
$14$ & $301$ & $227$ & $1.3260$ & $0.8209$ & & $47$ & $6252$ & $4868$ & $1.2843$ & $0.8257$ & $0.9260$\\
\hline
$15$ & $348$ & $272$ & $1.2794$ & $0.7987$ & & $48$ & $6610$ & $5128$ & $1.2890$ & $0.8282$ & $0.9331$\\
\hline
$16$ & $410$ & $318$ & $1.2893$ & $0.8008$ & $1.1230$ & $49$ & $7012$ & $5398$ & $1.2990$ & $0.8344$ & $0.9339$\\
\hline
$17$ & $488$ & $374$ & $1.3048$ & $0.8191$ & $1.1001$ & $50$ & $7447$ & $5675$ & $1.3122$ & $0.8425$ & $0.9347$\\
\hline
$18$ & $575$ & $433$ & $1.3279$ & $0.8366$ & $1.0708$ & $51$ & $7848$ & $5960$ & $1.3168$ & $0.8450$ & $0.9393$\\
\hline
$19$ & $656$ & $496$ & $1.3226$ & $0.8338$ & $1.0626$ & $52$ & $8262$ & $6248$ & $1.3223$ & $0.8474$ & $0.9458$\\
\hline
$20$ & $736$ & $563$ & $1.3073$ & $0.8229$ & $1.0700$ & $53$ & $8724$ & $6568$ & $1.3283$ & $0.8532$ & $0.9459$\\
\hline
$21$ & $836$ & $632$ & $1.3228$ & $0.8273$ & $1.0530$ & $54$ & $9221$ & $6890$ & $1.3383$ & $0.8606$ & $0.9460$\\
\hline
$22$ & $957$ & $716$ & $1.3366$ & $0.8431$ & $1.0360$ & $55$ & $9672$ & $7222$ & $1.3392$ & $0.8623$ & $0.9414$\\
\hline
$23$ & $1064$ & $804$ & $1.3234$ & $0.8388$ & $1.0091$ & $56$ & $10136$ & $7556$ & $1.3415$ & $0.8638$ & $0.9370$\\
\hline
$24$ & $1176$ & $895$ & $1.3140$ & $0.8335$ & $0.9831$ & $57$ & $10560$ & $7896$ & $1.3374$ & $0.8610$ & $0.9295$\\
\hline
$25$ & $1280$ & $992$ & $1.2903$ & $0.8192$ & $0.9472$ & $58$ & $10995$ & $8243$ & $1.3339$ & $0.8583$ & $0.9224$\\
\hline
$26$ & $1387$ & $1091$ & $1.2713$ & $0.8048$ & $0.9388$ & $59$ & $11400$ & $8604$ & $1.3250$ & $0.8527$ & $0.9123$\\
\hline
$27$ & $1480$ & $1204$ & $1.2292$ & $0.7814$ & $0.9483$ & $60$ & $11816$ & $8968$ & $1.3176$ & $0.8475$ & $0.9086$\\
\hline
$28$ & $1618$ & $1318$ & $1.2276$ & $0.7800$ & $0.9570$ & $61$ & $12196$ & $9354$ & $1.3038$ & $0.8393$ & $0.9098$\\
\hline
$29$ & $1796$ & $1442$ & $1.2455$ & $0.7931$ & $0.9486$ & $62$ & $12669$ & $9749$ & $1.2995$ & $0.8371$ & $0.9068$\\
\hline
$30$ & $1985$ & $1570$ & $1.2643$ & $0.8054$ & $0.9437$ & $63$ & $13220$ & $10146$ & $1.3030$ & $0.8393$ & $0.9051$\\
\hline
$31$ & $2148$ & $1704$ & $1.2606$ & $0.8029$ & $0.9464$ & $64$ & $13724$ & $10556$ & $1.3001$ & $0.8376$ &\\
\hline
$32$ & $2326$ & $1840$ & $1.2641$ & $0.8031$ & $0.9540$ & $65$ & $14256$ & $10972$ & $1.2993$ & $0.8370$ &\\
\hline
$33$ & $2532$ & $1996$ & $1.2685$ & $0.8095$ & $0.9489$ & & & & & &\\
\hline
\end{tabular}
\end{table}

\begin{table}
\centering
\small
\caption{$\phi$-values of optimal $n$-towns, calculated using Table~1 from \cite{dfr-psmd-11}.}\label{tbl-opt-phi}
\begin{tabular}{|r|r||r|r||r|r||r|r|}
\hline
$n$ & $\phi_{opt}(n)$ & $n$ & $\phi_{opt}(n)$ & $n$ & $\phi_{opt}(n)$ & $n$ & $\phi_{opt}(n)$\\
\hline \hline
$65$ & $0.644217$ & $69$ & $0.643676$ & $73$ & $0.645275$ & $77$ & $0.645053$\\
\hline
$66$ & $0.644281$ & $70$ & $0.644399$ & $74$ & $0.645136$ & $78$ & $0.645234$\\
\hline
$67$ & $0.644240$ & $71$ & $0.645067$ & $75$ & $0.645072$ & $79$ & $0.645524$\\
\hline
$68$ & $0.644104$ & $72$ & $0.645317$ & $76$ & $0.644715$ & $80$ & $0.645595$\\
\hline
\end{tabular}
\end{table}

A more refined analysis of $\Lambda(n)$ considers maximizing $\sum_i c_i^2 + \sum_j r_j^2$
all at once, instead of $\sum_i c_i^2$ and $\sum_j r_j^2$ separately, for a maximum
value of $n(2\sqrt{n}+5)+\frac{n^2}{2\sqrt{n}+5}$. For $n\geq 81$, this yields $2\Lambda(n)/n^{2.5}\leq \frac{2}{243}+\frac{5}{2187}+\frac{2}{621}=0.0137373...$
As the resulting competitive ratio of 1.7643 is only very slightly better, we omit further details.
If instead we rely on the unproven conjecture in \cite{dfr-psmd-11} that $\frac{2c_{{town}}}{n^{2.5}}\approx \psi-\frac{0.410}{n}$, 
we get $\phi\geq 0.6451$, which corresponds to experimental evidence; the resulting competitive factor is 1.7406.

\section{Lower Bounds}
We demonstrate that there are non-trivial lower bounds for a
competitive factor. We start by considering the discrete online scenario 
for towns.

\begin{theo}\label{theo-lower-bound}
No online strategy can guarantee a competitive factor below $\frac{64}{\sqrt{5}^5}=1.144866...$.
\end{theo}

\begin{figure}
  \centerline{\includegraphics[width=0.3\textwidth]{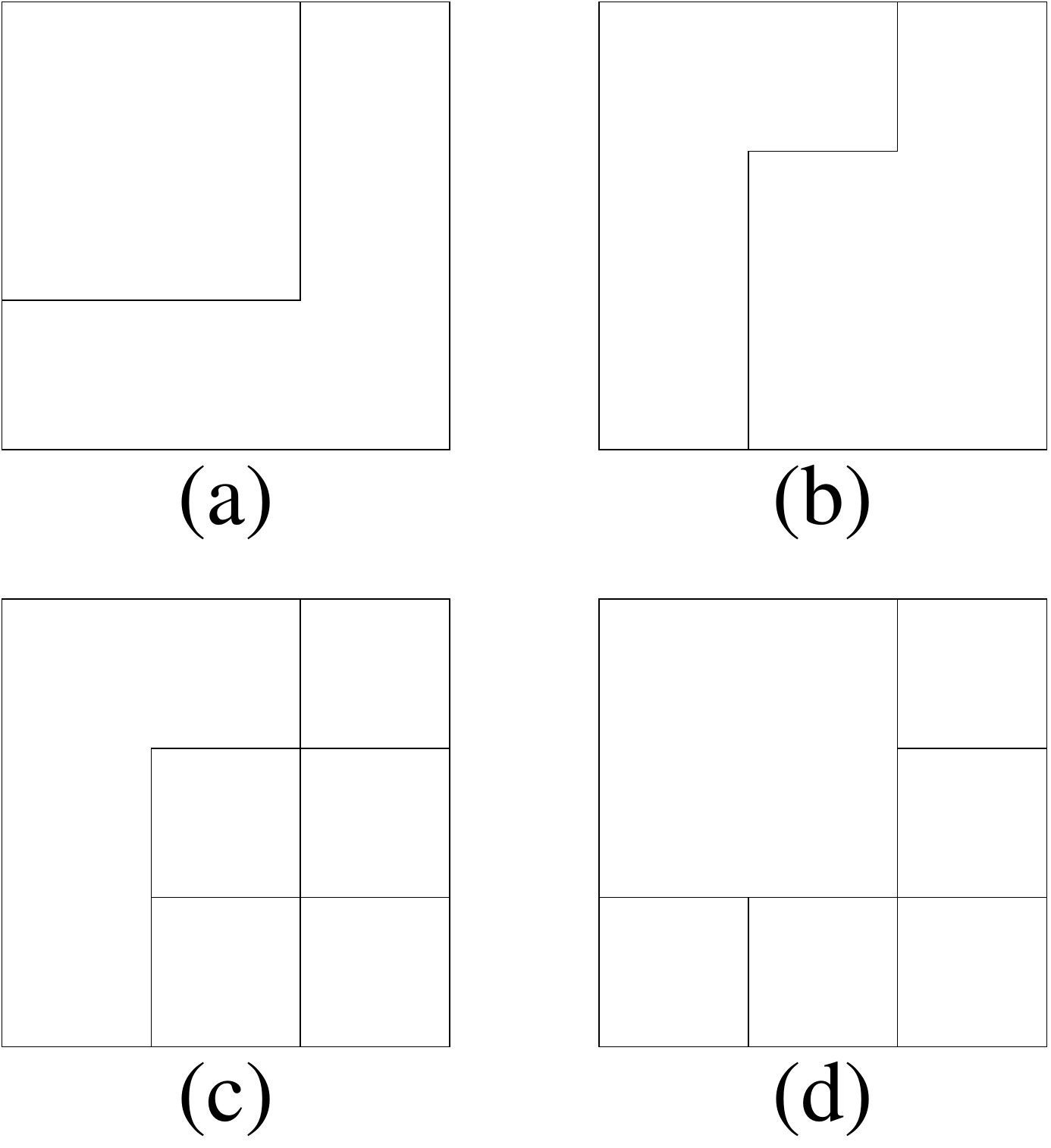}}
  \caption{The four cases considered in Theorem~\ref{theo-lower-bound}; the left column shows the choices by an algorithm, the right the corresponding
optimal choices for the ensung sequence.}\label{fig-lower-bound}
\end{figure}

{\bf Proof: } 
Consider a $3\times 3$ square, and let $n_1=4$; see Fig.~\ref{fig-lower-bound}. 
If (a) the strategy allocates a $2\times 2$ square (for a total distance of 8), then $n_2=5$, and the 
resulting L-shape has a total distance of 20 and a $\phi$-value of $40/5^{2.5}=0.715541...$
Allocating (b) the first town with an L-shape of total distance
10 results in $\phi=20/32=0.625$, and the second with a total distance of 16, or
$\phi=32/5^{2.5}=0.572433...$

If instead, (c) the first town is allocated different from a square, the
total distance is at least 10, and $\phi\geq 20/32$; then (d) $n_2=n_3=n_4=n_5=n_6=1$, and
an optimal strategy can allocate the first town as a 2x2 square, with $\phi=0.5$.
This bounds the competitive ratio, as claimed.
\hfill\qed

For the case of continuous allocations, we claim the following.

\begin{theo}\label{cont-lower-bound}
There is $\delta>0$, such that no 
online strategy can guarantee a competitive factor $1+\delta$.
\end{theo}

\begin{figure}
  \centerline{\includegraphics[width=0.5\textwidth]{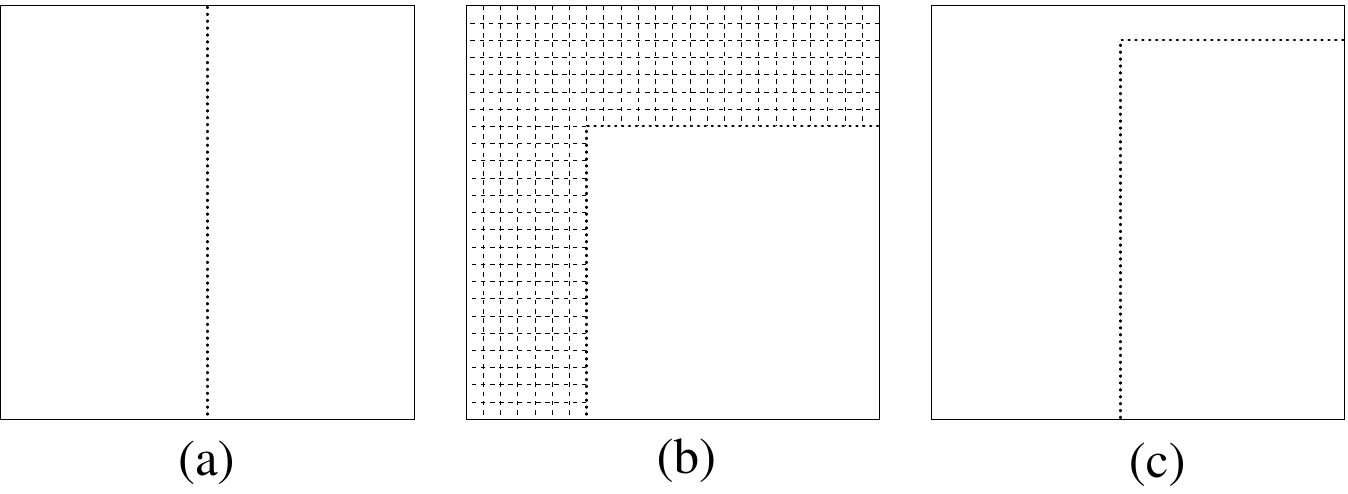}}
  \caption{The scenarios considered in Theorem~\ref{cont-lower-bound}, and a possible choice for the player.}\label{fig-lower-bound2}
\end{figure}

{\bf Proof: } 
Consider $n_1=1/2$, in combination with the two possible scenarios

{\bf (a)} $n_2=1/2$;

{\bf (b)} $n_2=n_3=\ldots=\varepsilon$.

In scenario (a), an adversary can assign two $(1\times 1/2)$-rectangles,
for a $\phi$-value of $0.707...$; in scenario (b), 
an adversary can assign all
shapes as squares, for a $\phi$-value of $0.666...$ If 
the player chooses a square size $\sqrt{2}/2$ first,
the adversary can choose scenario (a), causing the second allocation
to be in L-shape with $\phi$-value $\frac{2}{3}(7-4\sqrt{2})=0.895431...$,
as opposed to the optimal value of $0.707...$
If the player chooses a $(1\times 1/2)$-rectangle first, 
the adversary chooses scenario (b), for a ratio of $1.06066...$
The existence of the claimed lower bound follows from continuity,
as the $\phi$-value changes
continuously with continuous deformation of the involved shapes. \hfill\qed

The precise value arising from the scenarios in Theorem~\ref{cont-lower-bound}
is complicated. It can be obtained by computing the optimal
intermediate value for the player that allows him to protect against
both scenarios at once. For example, optimizing over the family of allocations 
shown in Figure~\ref{fig-lower-bound2} (c) yields a 
competitive ratio that is better than 1.06; however, the player may do
even better by using curved boundaries.
The involved computational effort for the resulting optimization
problem promises to be at least as complicated as computing
the ``optimal shapes of a city'', for which no closed-form solution
is known, see \cite{karp,bender}.

\section{Conclusions}
We have established a number of results for the online shape allocation problem.
In principle, further improvement could be achieved
by replacing the computational results for level 3 (i.e., $n=16,\ldots,64$) by level 4
(i.e., $n=65,\ldots,256$). (Conversely, a simplified
analysis with level 2, i.e., $n=4,\ldots,16$; yields a worse factor of 3.6525.) 
However, the highest known optimal $\phi$-values are for $n=80$, obtained 
by using the $O(n^{7.5})$ algorithm of \cite{dfr-psmd-11}. 
In any case, there is a threshold
of 1.3504 for Hilbert-based strategies, which we believe to be tight:
this is the ratio between the upper 
bound of 0.8768 for $n=14$ (and for $n=56,224,\ldots$)
and the asymptotic lower bound of 0.650245; because asymptotically, continuous and discrete case
converge, this also applies to the discrete case. 
Other open problems are to raise the lower bound of 1.144866 for the
discrete case, and establish definitive values for the continuous case.

As noted in Section~3, we can eliminate the assumption of all $n_i$ being multiples of
some $2^{-R}$, by making use of Lemma~\ref{lemma-wi-bound2}, and allocating a small round-off
fraction to a fractional pixel maintains the same bounds. However, the formal aspects
of describing the resulting allocation scheme become somewhat tedious and would
require more space than seems appropriate.

The offline problem is interesting in itself: for given $n_i$,
allocate disjoint regions of area $n_i$ in a square, such that
the maximum average Manhattan distance for each
shape is minimized. 
As mentioned, there is some indication that this is
an NP-hard problem; however, even relatively simple instances are prohibitively tricky
to solve to optimality, making it hard to give a formal proof. Clearly, our
online strategy provides a simple approximation algorithm; however, better factors
should be possible by exploiting the a-priori information of knowing all $n_i$, e.g.,
by sorting them appropriately.

\begin{figure}[h!]
  \centerline{\includegraphics[width=0.2\textwidth]{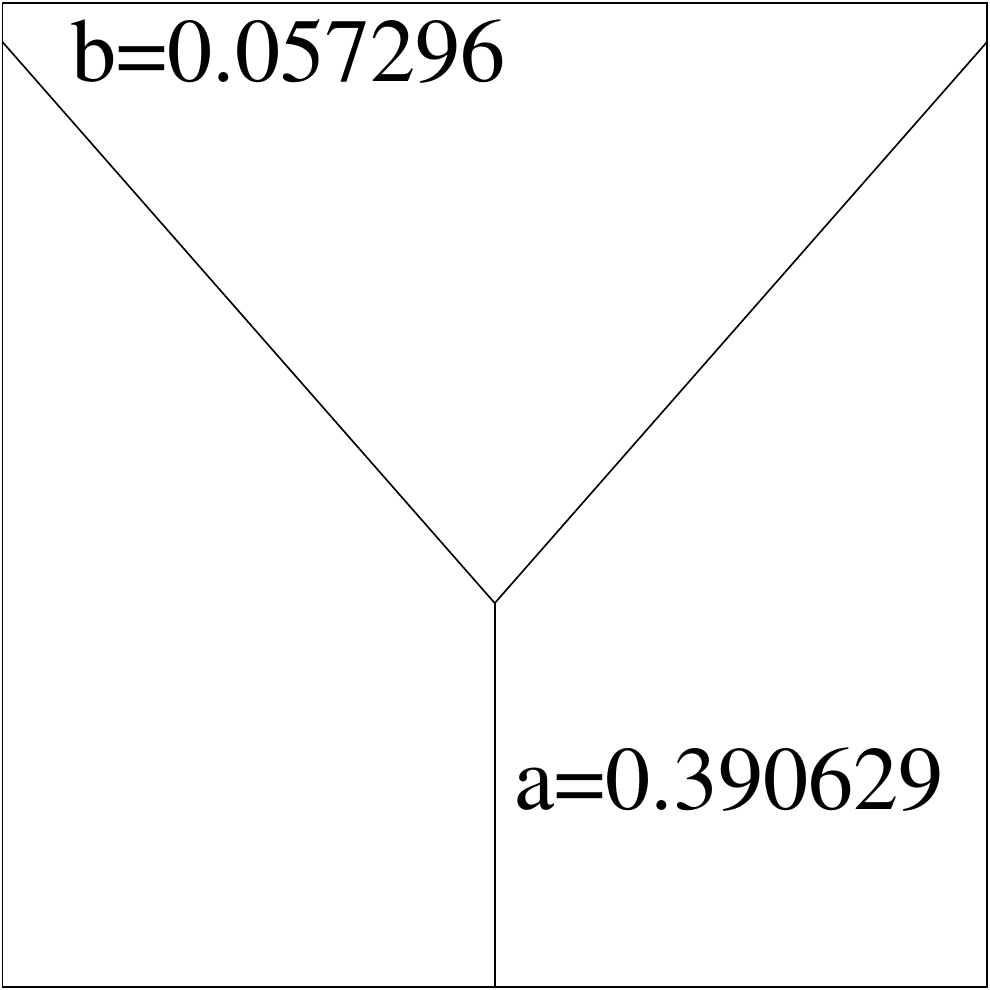}}
  \caption{A possible worst-case scenario for the offline problem.}\label{thirds}
\end{figure}

Another interesting open question for the offline scenario is the maximum optimal $\phi$-value
for any set $n_1,\ldots,n_i$.
A simple lower bound 
is $2/3=0.666...$, as that is the average distance of the whole square. A better lower bound is 
is provided by dividing 
the square into two or three equal-sized parts. For the case $n_1=n_2=1/2$,
we can use symmetry and convexity to argue that 
an optimum can be obtained by a vertical split, yielding
$\phi=\sqrt{2}/2=0.707$.
We believe the global worst case is attained for $n_1=n_2=n_3=1/3$.
Unfortunately, it is no longer possible to exploit only symmetry for arguing
global optimality.  
Figure~\ref{thirds} shows an allocation with $\phi=0.718736...$
for all three regions\footnote{More precisely, the involved values can be expressed as
${a=\frac{1}{108} \left(55-\frac{791}{\theta}+\theta\right)}$ and
${\phi}=\frac{\left(9602477-13416 \sqrt{585705}\right)\theta + 
 \left(202679+204 \sqrt{585705}\right)\psi^2 + 82133\theta^3}{77760 \sqrt{3} \theta}$ with
$\theta:=\left(-16253+36 \sqrt{585705}\right)^{1/3}$.}.
We conjecture that this is the best solution for
$n_1=n_2=n_3=1/3$, as well as the worst case for any optimal partition of the
unit square.




\subsubsection*{Acknowledgments.}
A short abstract based on preliminary results of this paper appears in the informal,
non-competitive Workshop EuroCG. (Standard disclaimer of that workshop:
``This is an extended abstract of a presentation given at EuroCG
2011. It has been made public for the benefit of the community
only and should be considered a preprint rather than a formally
reviewed paper. Thus, this work is expected to appear in a
conference with formal proceedings and/or in a journal.'')

We thank Bettina 
Speckmann for pointing out references \cite{w-nsfvs-05} and \cite{DBLP:journals/corr/abs-1012-1749}, and other colleagues for helpful hints to improve the presentation of this paper.

\thispagestyle{empty}
\small 
\bibliographystyle{abbrv}

\bibliography{lit}

\end{document}